# Using Blogs to Promote Writing Skill in ESL Classroom

MELOR MD. YUNUS, JULIAN LAU KIING TUAN & HADI SALEHI
Faculty of Education
Universiti Kebangsaan Malaysia
43600 UKM Bangi, Selangor
MALAYSIA
melor@ukm.my; julmark@gmail.com; hadisalehi1358@yahoo.com

*Abstract:-* This study provides details on the motivational factors for using blogs as an essential tool to promote students' writing skills in ESL classrooms. The study aims to discuss how using blogs may be integrated into classroom activities to promote students' writing skills as well as polishing their skills. It would also illustrate the features offered in blogs as well as the motivational essence that is attached to the blogs. To achieve the aim of the study, a semi-structured interview protocol was used to collect the required qualitative data. The findings of the study would serve as an insistent reminder that the blogs which have been clearly underlined in the curriculum should be re-orchestrated more effectively again by the teachers of English as a Second Language (ESL).

## 1 Introduction

The English language has been taught as a second language in all the Malaysian national schools so that it is very challenging to teach English in these schools. Moreover, in accordance with the globalization trend, English language has been instilled as a compulsory subject in curriculum for all the Malaysian national schools. English language learning in schools has always been linked with grammar, literature, exercises and drill activities that position teaching and learning a language as an educational activity that is related to the students' personal development in everyday lives. Now, technology has totally influenced pedagogy as the new ways of teaching [1-3]. With the growth of technology, Social Networking Sites (SNSs) have spread so rapidly and become the new phenomenon among the teenagers. Thus, SNSs are primarily concerned with people who already know each other, and use the Internet as one way of keeping their existing social connections alive, rather than for making new ones [4]. The most popular SNSs are often associated with the broader context of Web 2.0 technologies, which came to widespread prominence towards the end of 2004. According to Boyd and Ellison (2007), SNSs can be defined as web-based services that allow individuals to articulate a list of other users with whom they share a connection such as teacher and the students [5]. However, the growing popularity of blogs as educational tools can be explained by research in this area. Some researchers have claimed that students' writing skills improve when they blog [6]. A study conducted by Kavaliauskiene & Vaiciuniene (2006) indicated that the experience of writing on blogs (for an audience) provides opportunities to help students improve their knowledge of English [7]. Nadzrah (2007) also found that blogs let students compose writing with specific purposes that can encourage them to enhance their writing in the language constructively [8].

The focus of this study is limited to examining the perceptions of the selected lecturers who are familiar with the usage of blogs. They have been utilising blogs in their classes for some periods of time to actually find some solutions to improve their students' writing skills and encourage them to write at the same time. Moreover, it is assumed that blogs can encourage the undergraduates to write more consecutively in the future. Therefore, this study investigates the lecturers' perceptions of using blogs in ESL classrooms. It also reviews the advantages and benefits of using blogs in promoting students' writing skills in ESL classrooms. In order to facilitate the investigation regarding the effectiveness of blogs in promoting writing skills in ESL classrooms, the researchers formulated the following three research questions:

1. How do English lecturers perceive using blogs to promote students' writing skills?
2. What are the benefits and advantages of using blogs in encouraging ESL students to write?
3. What are the challenges encountered by the lecturers in using blogs to promote students' writing skills?

## 2 Literature Review

The term *blog* is a contraction of two words: *web* and *log*. Blogs are a tool for written communication and interaction. The term weblog refers to a personalized web page, kept by the author in reverse chronological diary form. As a "log on the web", it is kept first and foremost on the web, either on a static web page, or via a database-backed website, enabled through blogging software. As a "log of the web", it easily refers to other Internet locations via hyperlinks [9]. Blogs are personal online journals which have recently become a collaborative technology, and are regarded as a new way for people to express their thoughts in public [10]. This form of writing has become popular among Internet users. Most blog writers (bloggers) use this environment for self-expression and empowerment, as writing in blogs helps people become more thoughtful and critical in their writing [11].

Blogging software encourages frequent site updates with the content, or 'micro-content', as Alexander (2006) prefers to name it, being presented in reverse chronological order [12]. Hence, the micro-contents or posts are mainly composed of the blogger's own opinions and thoughts, showing why they are referred to as 'online diaries' [13]. Moreover, blogs can successfully promote self expression in a place where the L2 learner/blogger is developing deeply personalised content dealing with their language learning [14]. Being both potentially individualistic and collaborative, blogs can transcend linguistic barriers and may be used for language learning purposes, where bloggers become part of a discourse community in a complex multimodal setting, and learning together in collaborative spaces where people negotiate and construct meaning and texts [15-16].

Blogs can also be used to publish and exchange personal knowledge. The idea of using blogs in language learning is similar to the use of journal writing which is a way to help students explore and assimilate new ideas, create links between the familiar and unfamiliar, mull over possibilities, and explain them to others [10]. Every student can get the chance to express himself/herself and publish his/her thoughts. On the other hand, the teacher and other students can contribute their understanding on the publication. Hence, the students can strengthen their knowledge by constructing their own meanings in writing and interpret these meanings based upon their understanding and experiences. Blogs, therefore, should be used as they should be to promote writing skills among ESL learners.

## 3 Method

A qualitative study using semi-structured interview was used in exploring the ESL lecturers' perceptions of the effectiveness of blogs in promoting writing skills among ESL learners. The study was carried out in Universiti Kebangsaan Malaysia (UKM) in September 2010. Three major open-ended questions were used in the semi structured interview. The first major question and its sub-questions asked the participants to express their perceptions of using blogs in ESL classrooms. The second major question and its sub-questions were about how frequent blogs are used in teaching and learning, and what benefits using blogs has in encouraging ESL students to write. The third major question was about the challenges encountered by the teachers in using blogs to promote students' writing skills.

The interviews were carried out in English and all of them were recorded from the first second until the very last. At the end of each interview, the data were transcribed verbatim into a standard form which contained the interviewee's name (pseudonym), data, time and place of the interview and adequate space for transcribing the responses of the interviewee. Then, the data were analysed based upon the lecturers' gender, teaching experiences and competency in using blogs. A table was also created to include all the questions with regard to the lecturers' answers. It helped the researchers to compare all the given answers and identify the statements which are needed for this research.

## 4 Findings and Discussion

The findings are presented in three sections. Each section discusses the objectives of this research namely the lecturers' perceptions of using blogs, advantages and benefits of using blogs, as well as problems and challenges encountered by the lecturers in using blogs to promote students' writing skills. Four ESL lecturers (two males and two females) from Universiti Kebangsaan Malaysia were selected to participate in the interviews. Two interviewees had 7 - 9 years of experience in using blogs and the other two more had used blogs less than 3 years. Moreover, two interviewees were high users of blogs in ESL classrooms and the other two participants were modest users.

### 4.1 Perceptions on the Usage of Blogs

All the interviewed participants indicated that blogs allow the lecturers to integrate their site with other multimedia components. The interviews showed that the components and functions in blogs encourage ESL learners in promoting their writing skills. One of the interviewees stated that,

*in my tutorial classes, I had my students to choose on a topic they wanted to focus on and then write and elaborate about it. They have to include some components or functions which offered in Wordpress. They (my students) went along very well and I can see they move on week by week. Before this, during my previous semester, I asked them to write without referring to any graphics or anything. Less progress or efforts were put by my students. Some students even complained that this was just like some writing classrooms.*

The above comments suggest that the lecturers had eventually utilized blogs with one purpose that is to promote writing skills among the ESL students through exploiting all the possible functions and components provided. These approaches may take the same time to be prepared and explored by themselves or their students but worth a success.

### 4.2 Benefits of Using Blogs

There were lots of benefits and advantages cited by the lecturers during the interviews when blogs are used with their students. As indicated by one of the interviewees, blogs can help to seal the connection between the students and the lecturer. The lecturer added that,

*blogs help to gain free interaction so it's more non-formal, and there are no language barriers while writing in their blog. However, it won't be appropriate to simply say some improper commands of language to lecturer but it can be used in blog anytime. Students can also use blogs for education not totally for socialising. Students can ask about assignments and talk about lesson and etcetera.*

This statement was also supported by another quote from fellow lecturer saying that blogs help to,

*..communicate more in written English rather than doing it face to face. They are not shy to express themselves better.*

The above statements generally suggest that there are lots of benefits in using blogs such as in interaction and expressing themselves. This also supports that blogs provide a platform for individual expression and support reader commentary, critique, and interlink age as subsequent steps. In other words, blogs foreground the individual.

### 4.3 Challenges in Using Blogs

There are many challenges that hinder the lecturers from encouraging ESL learners to use blogs for improving their writing skills. The lack of time, students' lack of skills and less participation from the students are regarded as some of the existing challenges. As for the lack of time, one of the lecturers mentioned that:

*.....another problem will be time constraint whereby I need to spend quite some time to check their blogs. Students sometime also have less time to arrange*

*their works in blogs. Some even simply post useless things.*

However, as stated, it is not always about time but sometimes it is due to the lack of skill acquired by the students as highlighted by one of the lecturers:

*……I have Master students from Middle East, and they do not know how to use mouse, because they have not used computer at home, and they don't even know how to surf the net. Even they don't have an email. So, these are the basic requirements for my course. If they don't have these requirements, then I have to make sure they settle this level of literacy before enrolling in my course.*

When it is not about time and skill, sometimes the students are too passive in involving themselves through blogs. One of the interviewees mentioned that:

*…..some students don't participate, or they participate in the last minute. Even if they participate, they just give you some lines. They just write a few lines and you need to encourage them.*

Therefore, although blogs are very useful, there are still some challenges which need to be tackled to ensure their effectiveness.

## 5 Conclusions

Basically, this study can be considered as a success because all of the objectives were achieved and all the research questions were answered. All the interviewed lecturers agreed that blog is a very useful tool for promoting writing skills among the ESL learners. Although some problems may occur and hinder the lecturers to use blogs, some solutions have been set to abolish these weaknesses. Therefore, the ESL learners can be motivated to improve their writing skills through using blogs.

*Reference:*
[1] Melor Md. Yunus. Language Learning via ICT: Uses, Challenges and Issues, *WSEAS Transactions on Information Science and Applications,* Vol. 6, No. 9, 2009, pp. 1453-1467.
[2] Maimun Aqsha Lubis, Mohamed Amin Embi, Melor Md Yunus, Ismail Wekke, & Nor Azah Nordin. The Application of Multicultural Education and Applying ICT on Pesantren in South Sulawesi, Indonesia. *WSEAS Transactions on Information Science and Applications*, Vol. 6, No. (8), 2009, pp. 1401 1411.
[3] Melor Md Yunus, Salehi, H., & Chenzi, C. Integrating Social Networking Tools into ESL Writing Classroom: Strengths and Weaknesses. *English Language Teaching,* Vol. 5, No. 8, 2012, pp. 42-48. doi:10.5539/elt.v5n8p42
[4] Harrison, R., & Thomas, M. Identity in Online Communities: Social Networking Sites and Language Learning. *International Journal of Emerging Technologies & Society,* Vol. 7, No. 2, pp. 109 –124. 2009.
[5] Boyd, D., & Ellison N. Social Network Sites: Definition, History and Scholarship. *Journal of Computer-Mediated Communication*, Vol. 13, No. 1, 2007, pp. 210-230.
[6] Hall, H., & Davison, B. Social Software as Support in Hybrid Learning Environments: The Value of the Blog as a Tool for Reflective Learning and Peer Support. *Library and Information Science Research,* Vol. 29, No. 2, 2007, pp. 163-187.
[7] Kavaliauskienė, G., & Vaičiūnienė, V. Communication Interaction Using Information and Communication Technology. *Studies about Languages (Kalbų Studijos),* 8, 2006, pp. 8894-8899.
[8] Nadzrah Abu Bakar. English Language Activities in Computer-based Learning Environment: A Case Study in ESL Malaysian Classroom, *GEMA Online Journal of Language Studies,* Vol. 7, No. 1, 2007, pp. 33-49.
[9] Simsek, O. The Effect of Weblog Integrated Writing Instruction on Primary School Students Writing Performance. *International Journal of Instruction,* Vol. 2, No. 2. 2009.
[10] Nadzrah Abu Bakar. E-Learning Environment: Blogging as a Platform for Language Learning. *The European*

*Journal of Social Science*, Vol. 9, 2009, pp 594-604.

[11] Blood, R. *Weblogs: A History and Perspective*, In Editors of Perseus Publishing (Eds.). 2002.

[12] Alexander, B. Web 2.0: A new wave of Innovation for Teaching and Learning? *EDUCAUSE Review*, March/April, 2006, pp. 33-44.

[13] Hourigan, T., & Murray, L. Using Blogs to Help Language Students to Develop Reflective Learning Strategies: Towards a Pedagogical Framework. *Australasian Journal of Educational Technology,* Vol. 26, No. 2, 2010, pp. 209-225.

[14] Murray, L., & Hourigan, T. Blogs for Specific Purposes: Expressivist or Socio-cognitivist Approach? *ReCALL*, Vol. 20, No. 1, 2008, pp. 83-98.

[15] Richardson, W. Blogs, Wikis, Podcasts, and Other Powerful Web Tools for Classrooms. Thousand Oaks, CA: Corwin Press, 2006.

[16] Raith, T., The Use of Weblogs in Language Education. In M. Thomas (Ed), Handbook of Research on Web 2.0 and Second Language Learning. IGI Global, 2009, pp. 274-291.